\documentclass[%
superscriptaddress,
preprint,
 amsmath,amssymb,
 aps,
 pre,
]{revtex4-1}

\usepackage{graphicx}
\usepackage{dcolumn}
\usepackage{bm}
\usepackage{xcolor}

\usepackage{algorithm}
\usepackage{algorithmic}
\usepackage{pifont}

\newlength\myindent
\begin{document}


\title{Heterogeneous partition of cellular blood-borne nanoparticles through microvascular bifurcations}

\author{Zixiang L. Liu}
\email{zxliu@gatech.edu}
\affiliation{%
 The George W. Woodruff School of Mechanical Engineering, Georgia Institute of Technology, Atlanta, GA, 30332 USA}
\affiliation{ The Parker H. Petit Institute for Bioengineering and Bioscience, Georgia Institute of Technology, Atlanta, GA, 30332 USA
}%
\author{Jonathan R. Clausen}
\affiliation{
 Thermal and Fluid Processes, Sandia National Laboratories, Albuquerque, NM, 87185, USA
}%
\author{Justin L. Wagner}
\affiliation{
 Aerosciences Department, Sandia National Laboratories, Albuquerque, NM, 87185, USA
}%
\author{Kimberly S. Butler}
\affiliation{
 Molecular and Microbiology, Sandia National Laboratories, Albuquerque, NM, 87185, USA
}%
\author{Dan S. Bolintineanu}
\affiliation{
 Fluid and Reactive Processes, Sandia National Laboratories, Albuquerque, NM, 87185, USA
}%
\author{Jeremy B. Lechman}
\affiliation{
 Fluid and Reactive Processes, Sandia National Laboratories, Albuquerque, NM, 87185, USA
}%
\author{Rekha R. Rao}
\email{rrrao@sandia.gov}
\affiliation{
Fluid and Reactive Processes, Sandia National Laboratories, Albuquerque, NM, 87185, USA
}%
\author{Cyrus K. Aidun}
\email{cyrus.aidun@me.gatech.edu}
\affiliation{%
 The George W. Woodruff School of Mechanical Engineering, Georgia Institute of Technology, Atlanta, GA, 30332 USA}
\affiliation{ The Parker H. Petit Institute for Bioengineering and Bioscience, Georgia Institute of Technology, Atlanta, GA, 30332 USA
}%

\date{\today}

\newpage

\begin{abstract}
Blood flowing through microvascular bifurcations has been an active research topic for many decades, while the partitioning pattern of nanoscale solutes in the blood remains relatively unexplored. Here, we demonstrate a multiscale computational framework for direct numerical simulation of the nanoparticle (NP) partitioning through physiologically-relevant vascular bifurcations in the presence of red blood cells (RBCs). The computational framework is established by embedding a newly-developed particulate suspension inflow/outflow boundary condition into a multiscale blood flow solver. The computational framework is verified by recovering a tubular blood flow without a bifurcation and validated against the experimental measurement of an intravital bifurcation flow. The classic Zweifach-Fung (ZF) effect is shown to be well captured by the method. Moreover, we observe that NPs exhibit a ZF-like heterogeneous partition in response to the heterogeneous partition of the RBC phase. The NP partitioning prioritizes the high-flow-rate daughter branch except for extreme (large or small) suspension flow partition ratios under which the complete phase separation tends to occur. By analyzing the flow field and the particle trajectories, we show that the ZF-like heterogeneity in NP partition can be explained by the RBC-entrainment effect caused by the deviation of the flow separatrix preceded by the tank-treading of RBCs near the bifurcation junction. The recovery of homogeneity in the NP partition under extreme flow partition ratios is due to the plasma skimming of NPs in the cell free layer. These findings, based on the multiscale computational framework, provide biophysical insights to the heterogeneous distribution of NPs in microvascular beds that are observed pathophysiologically.
\end{abstract}

\maketitle

\section{Introduction}\label{sec:intro}
Blood is a biological fluid that conveys nutrients and wastes throughout the body. The primary constituents of blood are red blood cells (RBCs) that comprise $\sim$45\% of the systemic blood volume and plasma that contains a variety of biomolecules including clotting factors and nutrients, etc. The exchange of materials between the blood and tissues primarily occurs in the microvasculatures where blood vessels feature complex tree-like bifurcating structures. Such microvascular bifurcations allow the blood flow to be partitioned and disseminated to the bulk of tissues for efficient mass exchanges~\cite{sherwood2015human}. 

Given its significance in physiology and fluid mechanics, blood flowing through microvascular bifurcations has been a topic of research for many decades~\cite{pries1990blood,pries2008blood,secomb2017blood}. Among all the studies, substantial efforts have been dedicated to the understanding of the pattern and biomechanistic mechanisms of RBC partitioning through single or multi-generation bifurcations using in vivo~\cite{svanes1968,pries1989red}, in vitro~\cite{bugliarello1964,fung1973} and theoretical~\cite{yan1991three,pries1990blood,Bagchi2018POF} approaches. It has been well accepted that the RBC volume concentration (i.e., hematocrit) in the two daughter branches are oftentimes not equally partitioned due to the particulate nature of the blood cells. Rather, the hematocrit in the high-flow-rate daughter branch often gets elevated while the other daughter branch receives a diluted or even depleted hematocrit level. This heterogeneous partition of RBCs, often referred to as the Zweifach–Fung (ZF) effect~\cite{svanes1968,fung1973}, is physiologically essential since it biomechanistically explains the reduction and heterogeneous distribution of the hematocrit in microvasculatures that has been observed physiologically~\cite{johnson1971red,kanzow1982analysis}. Utilizing the ZF effect, novel microfluidic devices have also been developed for cell or bioprotein seperation and other relevant biomedical applications~\cite{yang2006microfluidic,fan2008integrated}.

Besides the large body of research focusing on the RBC partition through microvascular bifurcations, a few studies investigate the distribution and adhesion of other blood species through bifurcating structures.
The effect of size and shape of microscale particles, mimicking white blood cells, on their adhesion propensity at bifurcations has been studied in vitro using synthetic microvascular networks~\cite{tousi2010preferential,doshi2010flow}. The model platelet distribution in blood flow through idealized, vascular bifurcations or confluences has been simulated directly, where the platelets shows anti-margination behaviors at the downstream of a confluence~\cite{bacher2018antimargination}. The distribution and adhesion of nanoparticles (NPs) in vascular bifurcations has been interrogated in silico without explicit consideration of RBCs~\cite{tan2013influence,sohrabi2017nanoparticle}.

Notwithstanding current understanding of the RBC partitioning characteristics through microvascular bifurcations, there is clearly a lack of knowledge regarding the partitioning behaviors of nanoscale solutes through capillary beds subject to the influence of RBCs, partly because of the difficulty imposed by the multiscale nature of the problem~\cite{Liu2018a}. Nonetheless, studying the partition of nanoscale solutes through microvascular bifurcations holds considerable significance biophysically and clinically. \textcolor{black}{For example, the undesired heterogeneous distribution of nanomedicine in tumor microvasculature remains to be a major hurdle that limits the overall therapeutic efficiency~\cite{jain2010delivering}. Understanding the partitioning of solutes through microvascular bifurcations may provide possible mechanisms that could be applied to regulate the heterogeneity of solute distribution in microcirculation. Accordingly, novel design of NP-based drugs with improved drug delivery efficacy could be inspired~\cite{wilhelm2016analysis}}.

The goal of this paper is to develop a multiscale computational framework for direct numerical simulation of NP partition through microvascular bifurcations, while quantitatively elucidating the NP partitioning in the presence of RBCs. Towards that end, an general index-based manipulation strategy for the particle inflow/outflow boundary conditions embedded in a verified and validated multiscale blood flow solver~\cite{Liu2018a,Liu2018b,Liu2019a} has been developed and validated. Using the multiscale computational framework, the partition of the blood-borne NPs through microvascular bifurcations in response to the ZF effect is found to be heterogeneous as well. The heterogeneity in the NP partition is conditional depending on the ratio of the flow rates between the daughter branches. The physical mechanisms responsible for the heterogeneity in the NP partition is found to be associated with the flow separatrix deviation at the bifurcation junction caused by the RBC motion.

The remainder of the paper is organized as follows. In \S~\ref{sec:multiblood}, the detailed computational methods are presented. In \S~\ref{sec:exp}, the in vivo experimental setup is described. The validity of the computational methods is demonstrated in \S~\ref{sec:vef1} and \S~\ref{sec:vef2}. \S~\ref{sec:actualres} discusses the pattern and the mechanisms for NP partitioning through microvascular bifurcations. In \S~\ref{sec:conclusion}, we summarize the study.

\section{Computational methods}\label{sec:multiblood}
\begin{figure}[H]
\centering
\includegraphics[width=0.9\textwidth]{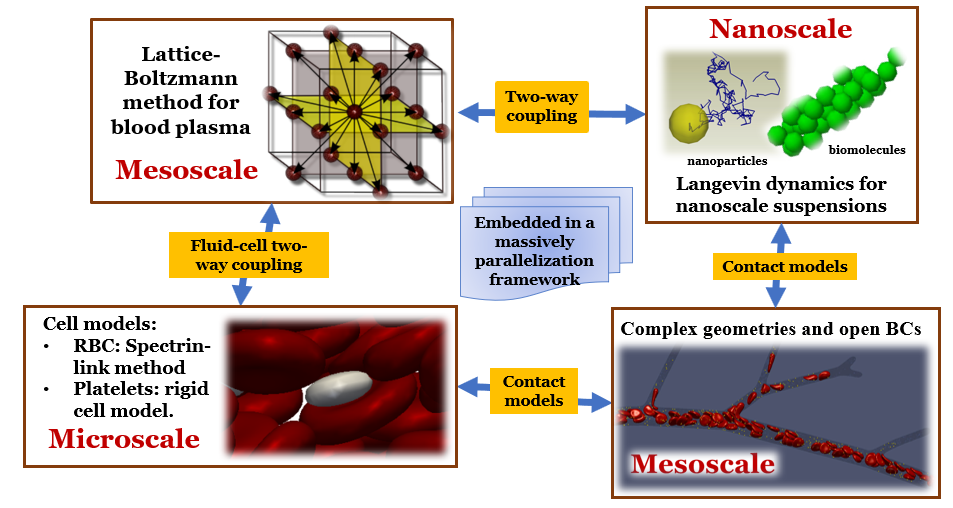}
\caption{The multiscale computational framework for simulating blood-borne nanoparticle partitioning through microvasculatures. The method directly and concurrently resolves the dynamics of microscale cells and nanoscale particles and molecules suspended in blood plasma through complex microvascular networks, where open boundary conditions (BCs) are prescribed.}
  \label{fig:frame}
\end{figure}

The computational framework proposed for this study is a 3D lattice-Boltzmann (LB) based multiscale complex blood flow solver~\cite{Aidun1998,Aidun2010,Reasor2012,Liu2018a,Liu2018b}, augmented with a particulate inflow and outflow boundary condition. The LB method is a well-established numerical model for hydrodynamics and proves to be a highly scalable method for direct numerical simulation (DNS) of dense particle suspensions~\cite{ClausenCPC2010,Aidun2010}. Modeling of the RBC dynamics and deformation is via a coupled spectrin-link (SL)/LB method~\citep{Reasor2012,Reasor2012,ReasorJFM2013}. The NP suspension dynamics are resolved via a two-way coupled lattice-Boltzmann Langevin-dynamics (LB-LD) method with both particle Brownian motion and long-range hydrodynamic interactions (HI) directly resolved and validated~\citep{Liu2018a,Liu2018b}. The solver has been applied to a variety of problems pertaining to particle and biomolecule transport in blood flow\citep{ReasorJFM2013,ReasorABE2013,MM2016,Ahmed2018,Liu2018a,Liu2019a,Liu2019b}. The newly developed open boundary conditions (BCs) are especially suitable for complex geometries such as microvasculatures. All modules are two-way coupled and embedded in a massively parallelized framework, as shown in Fig~\ref{fig:frame}. 

\subsection{Lattice-Boltzmann method for fluid motion}\label{sec:lb}
The fluid phase, including the blood plasma and the hemoglobin within RBCs, is simulated using the LB method particularly designed for suspension hydrodynamics~\cite{Aidun1995,Aidun1998,Aidun2010}. The method solves the discretized Boltzmann transport equation in velocity space through a streaming-and-collision procedure. In the streaming step, the fictitious fluid particles propagate along discrete velocity vectors in a lattice space. In the collision step, the local particle distribution function (PDF) relaxes into a local `Maxwellian' equilibrium distribution. The collision term is linearized based on the Bhatnagar, Gross, and Krook (BGK) operator \citep{Bhatnagarp} for simplicity. The governing equation of the PDF reads
\begin{equation}
\begin{split}
f_i(\boldsymbol{r}+\Delta t \boldsymbol{e}_i, t+\Delta t) = f_i(\boldsymbol{r}, t) - \frac{\Delta t}{\tau}[f_i(\boldsymbol{r}, t) - f_i^{(eq)}(\boldsymbol{r}, t)] + f_i^S(\boldsymbol{r}, t),
  \label{eqn:lb1}
\end{split}
\end{equation}
where $f_i$ is the fluid PDF, $f_i^{(eq)}$ is the equilibrium PDF, $r$ is the lattice site, $e_i$ is the discrete lattice velocity, $t$ is time, $\tau$ is the single relaxation time and \textcolor{black}{$f_i^S$ is a forcing source term introduced to account for the discrete external force effect that leads to corrections to the macroscopic variables.} This method has a pseudo speed of sound, $c_s=\Delta r/(\sqrt{3} \Delta t)$, and a fluid kinematic viscosity, $\nu$=$(\tau - \Delta t/2) c_s^2$, where $\Delta t$ is the time step and $\Delta r$ is the unit lattice distance. The positivity of $\nu$ requires $\tau$$>$$\Delta t/2$. In the LB method, time and space are typically normalized by $\Delta t$ and $\Delta r$, respectively, such that $\Delta t_{LB}$=$\Delta r_{LB}$=1 are employed to advance equation \ref{eqn:lb1}. In the near incompressible limit (i.e., the Mach number, $Ma$=$u/c_s$$\ll$1), the LB equation recovers the Navier-Stokes equation \citep{Junk2003} with the equilibrium PDF given in terms of local macroscopic variables as
\begin{equation}
  f_i^{eq)}(\boldsymbol{r}, t) = \omega_i \rho [1 +  \frac{1}{c_s^2}(\boldsymbol{e}_i \cdot \boldsymbol{u}) + \frac{1}{2c_s^4}(\boldsymbol{e}_i \cdot \boldsymbol{u})^2 - \frac{1}{2c_s^2}(\boldsymbol{u} \cdot \boldsymbol{u})],
  \label{eqn:lb2}
\end{equation}
where $\omega_i$ denotes the set of lattice weights defined by the LB stencil in use. \textcolor{black}{The macroscopic properties such as the fluid density, $\rho$, velocity, $\boldsymbol{u}$, and pressure, $p$, can be recovered via moments of the PDFs as
$\rho=\sum_{i=1}^Q f_i(\boldsymbol{r},t)$, 
$\rho\boldsymbol{u}=\sum_{i=1}^Q f_i(\boldsymbol{r},t) \boldsymbol{e}_i$ and $\rho\boldsymbol{u}\boldsymbol{u} = \sum_{i=1}^Q f_i(\boldsymbol{r},t) \boldsymbol{e}_i\boldsymbol{e}_i - p\mathbf{I}$,
respectively. Here, $\mathbf{I}$ is the identity tensor and pressure can be related to density and the speed of sound through $p$=$\rho c_s^2$. For the D3Q19 stencil adopted in the current study, $Q$ is equal to 19. Along the rest, non-diagonal, and diagonal lattice directions, $\omega_i$ is equal to 1/3, 1/18, and 1/36, and $|\boldsymbol{e}_i|$ is equal to 0, $\Delta r/\Delta t$, and $\sqrt{2}(\Delta r/\Delta t)$, correspondingly.}

\subsection{Spectrin-link method for RBC membrane dynamics}\label{sec:sl}
RBC deformation and dynamics are through a coarse-grained spectrin-link (SL) method \citep{FedosovBJ2010,Pivkin2008} coupled with the LB method. This method has been previously validated against experimental results and is proved to be a superior method compared to finite-element method~\citep{Reasor2012}. 
In the SL model, the RBC membrane is modeled as a triangulated network with a collection of vertices mimicking actin vertex coordinates, denoted by $\{\boldsymbol{x}_n,\ n$$\in$$1,...,N\}$. The Helmholtz free energy of the network system, $E(\boldsymbol{x}_n)$, including in-plane, bending, volume and surface area energy components \citep{Dao2006}, is given by 
\begin{equation}
  E(\boldsymbol{x}_n) = 
  E_{IP} + E_{B} + E_{\Omega} + E_{A},
  \label{eqn:sl1}
\end{equation}
where the in-plane energy, $E_{IP}$, prescribes the membrane shear modulus through a worm-like chain (WLC) potential \citep{Bustamante2003} coupled with a hydrostatic component \citep{FedosovBJ2010}. The bending energy, $E_B$, specifies the membrane bending stiffness, which is essential in characterizing the equilibrium RBC biconcave morphology \citep{Dao2006,FedosovBJ2010}. The constraint energies in volume, $E_{\Omega}$, and area, $E_A$, ensure the conservation of the RBC volume and area, respectively, in the presence of external forces. 

The dynamics of each vertex advance according to the Newton$\text{'}$s equations of motion,
\begin{equation}
  \frac{d\boldsymbol{x}_n}{dt}=\boldsymbol{v}_n;\ \ M\frac{d\boldsymbol{v}_n}{dt}=\mathbf{f}_n^{SL}+\mathbf{f}_n^{LB}+\mathbf{f}_n^{CC}
  \label{eqn:sl2}
\end{equation}
where $\boldsymbol{v}_n$ is the velocity of the vertex, $n$, and $M$ is the fictitious mass equal to the total mass of the cell divided by the number of vertices, $N$. \textcolor{black}{The number of vertices used to discretize the RBC membrane is $N$=613. This corresponds to a RBC link length of 1.5 lattice units ($\sim$450 $nm$), which has shown to be a cost effective resolution to capture both the single RBC dynamics \citep{Reasor2012} and concentrated RBC suspension rheology \citep{ReasorJFM2013} when coupled with the LB method}. $\mathbf{f}_n^{LB}$ specifies the forces on the vertex due to the fluid-solid coupling. $\mathbf{f}_n^{CC}$ are the forces due to cell-cell interactions. The forces due to the Helmholtz free energy based on the SL model is determined by
\begin{equation}
  \mathbf{f}_n^{SL} = -\frac{\partial E(\boldsymbol{x}_n)}{\partial \boldsymbol{x}_n}.
  \label{eqn:sl3}
\end{equation}
The SL method is solved by integrating equations \ref{eqn:sl2} at each LB time step using a first-order-accurate forward Euler scheme in consistency with the LB evolution equation to avoid excessive computational expense \citep{Reasor2012,Liu2018a}.

\subsection{Langevin dynamics for nanoparticle suspension dynamics}\label{sec:ld}
The nanoscale particle suspensions are resolved through a two-way coupled LB-LD method \citep{Liu2018a,Liu2018b}. \textcolor{black}{This method introduces the thermal fluctuation from the Brownian particles and performs momentum exchange between the particle phase and non-fluctuating LB fluid. Different from approaches coupled with fluctuating LB method~\cite{Ahlrichs1998,Ahlrichs1999}, the current method captures the theoretical Brownian diffusivity and the long-range many-body HI simultaneously without introducing empirical rescaling~\cite{Liu2018b}.} The dynamics of the LD particles is described through Langevin equation (LE),
\begin{equation}
  m \frac{d \boldsymbol{u}_p}{dt} = \boldsymbol{C}_p + \boldsymbol{F}_p + \boldsymbol{S}_p,
  \label{eqn:ld1}
\end{equation}
where $m$ is the mass of a single particle. The conservative force, $\boldsymbol{C}_p$, is determined by calculating the directional derivatives of the total potential energy $U_{total}$ as
\begin{equation}
  \boldsymbol{C}_p = - \frac{dU_{total}}{d\boldsymbol{r}_p},
  \label{eqn:ld2}
\end{equation}
where the $U_{total}$ specifies the interparticle and particle-surface interaction forces. The frictional force, $\boldsymbol{F}_p$, is assumed to be proportional to the relative velocity of the particle with respect to the local viscous fluid velocity \citep{Ahlrichs1998,Ahlrichs1999},
\begin{equation}
  \boldsymbol{F}_p = -\zeta[\boldsymbol{u}_p(t)-\boldsymbol{u}(\boldsymbol{r}_p,t)],
  \label{eqn:ld3}
\end{equation}
where $\boldsymbol{u}_p$ denotes the particle velocity, and $\boldsymbol{u}(\boldsymbol{r}_p,t)$ is the interpolated LB fluid velocity at the center of the particle. The friction coefficient, $\zeta$, is determined by the Stokes’ drag law, $\zeta=3\pi\mu d_p$, where $\mu$ is the dynamic viscosity of the suspending fluid. The stochastic force, $\boldsymbol{S}_p$, satisfies the fluctuation-dissipation theorem (FDT) \citep{Kubo1966} by
\begin{equation}
\langle S_{p,i}^{\alpha} (t)\rangle = 0;\ \  
\langle S_{p,i}^{\alpha} (t)S_{p,j}^{\beta} (t)\rangle = 2k_BT\zeta \delta_{ij}\delta_{\alpha\beta}\delta(t-t'),
\label{eqn:ld6}
\end{equation}
where $i,j\in \{ x,y,z\}$, $\alpha$ and $\beta$ run through all the particle indices, $\delta_{ij}$ and $\delta_{\alpha\beta}$ are Kronecker deltas, $\delta(t-t')$ is the Dirac-delta function, $k_B$ is the Boltzmann constant and $T$ is the absolute temperature of the suspending fluid. The angle brackets denote the ensemble average over all the realizations of the random variables. Since only the time scales equal to and greater than the Brownian diffusion time scale is of interest, this study solves the over-damped discretized LE as suggested in \cite{Liu2018a,Liu2018b}.

\subsection{Open boundary strategies for particulate suspension inflow and outflow}\label{sec:PSIO}
\begin{figure}[H]
	\centering
	\includegraphics[width=0.85\columnwidth]{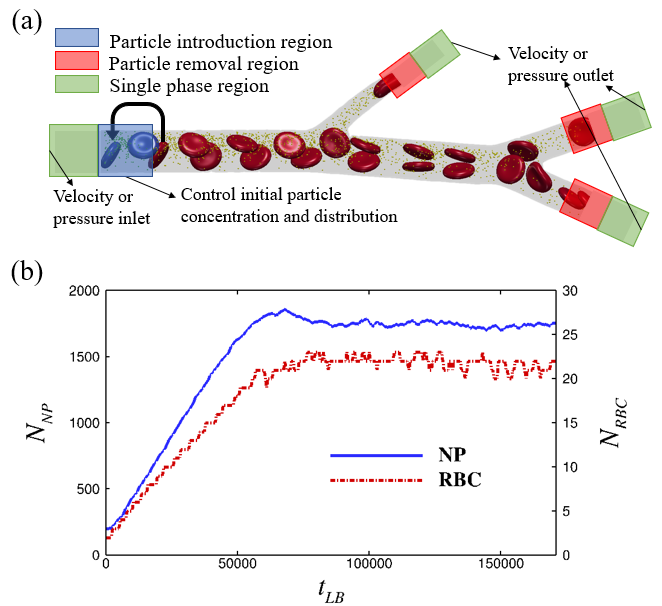}
	\caption{Demonstration of the particulate suspension inflow and outflow boundary condition (PSIO-BC). The schematic illustration of the PSIO-BC applied to a NP-RBC bidisperse suspension flowing through a two-generation microvascular bifurcation (a). The time course of the particle number simulated in the bifurcation (b), where at equilibrium the particle number becomes quasi-conserved.
	}
	\label{fig:PSIOBC}
\end{figure}

Periodic boundary condition (P-BC), due to its computational efficiency, has been widely used for simulations of blood flows through straight tubes/channels~\cite{Zhao2011,MM2016,Liu2018a,Liu2019b} or artificial bifurcation structures~\cite{balogh2017computational,Bagchi2018POF}. Physiologically, microvascular bifurcations often involve one inlet but multiple outlets where flow conditions are non-periodic. To overcome the methodological limitation, several open boundary strategies have been developed for inflow and outflow of particle or cell suspensions in the context of dissipative particle dynamics (DPD)~\cite{tarksalooyeh2018inflow,li2020parallel} or immersed-boundary (IB)/LB method~\cite{lykov2015inflow}. Here we propose a general particulate suspension inflow and outflow boundary condition (PSIO-BC), which in principle is applicable to any index-based particle suspension methods. Applying the PSIO-BC to both the LB-SL method~\cite{Reasor2012} and the LB-LD method~\cite{Liu2018b}, we simulate the RBC-NP suspension through a model microvascular bifurcation (Fig~\ref{fig:PSIOBC}a) where the number of particles initially increases and eventually plateaus when the system reaches equilibrium (Fig~\ref{fig:PSIOBC}b).

\subsubsection{Fluid phase}
The Dirichlet BCs are applied to prescribe the velocity or pressure of the flow field at the inlet and outlet. In the context of LB method, we employ the regularization BC procedure~\cite{latt2006lattice,latt2008straight} to construct the incoming unknown PDFs based on prescribed macroscopic properties at the inlet and outlet as
\begin{equation}
  f_i^R = f_i^{(eq)} + f_i^{(neq)} = f_i^{(eq)}(\rho,\boldsymbol{u}) + \frac{\omega_i}{2c_s^4}\boldsymbol{Q}_i:\boldsymbol{\Pi}^{(1)},
 \label{eqn:RegBC}
\end{equation}
where the tensor $\boldsymbol{Q}_i$ is defined as $\boldsymbol{Q}_i=\boldsymbol{e}_i\boldsymbol{e}_i-c_s\boldsymbol{I}$ and the stress tensor $\boldsymbol{\Pi}^{(1)}$ is evaluated based on the non-equilibrium components of the outward-pointing PDFs (in the opposite directions of the unknown PDFs)~\cite{latt2008straight}. 
Single phase regions are arranged to the upstream of the particle introduction region as well as to the downstream of the particle removal regions, as shown in Fig~\ref{fig:PSIOBC}a. This way, the velocity or pressure can be prescribed within the single phase region such that simple analytical solutions can be used. The aspect ratio of the single phase regions is chosen to be roughly one to allow smooth transitions of the properties from the single-phase to multi-phase regions or vice versa.

\subsubsection{Particle phase}
The introduction of particles at inlet is realized by inserting a particle introduction section downstream proximal to the single phase region, as depicted in Fig~\ref{fig:PSIOBC}a. Inside the seeding section, the number of particles is conserved to control the particle concentration and number flow rate introduced to the downstream. Similar to other approaches~\cite{lykov2015inflow,tarksalooyeh2018inflow,li2020parallel}, \textcolor{black}{particles in this section periodically leave and re-enter the domain, while at the exit new particles are introduced to the downstream by duplicating the particle (and its properties such as mesh coordinates, velocity and mechanical properties, etc) that are wrapped to the entrance (upstream end) of the particle introduction region. Besides, random motions can be introduced to the particles within the introduction region to reduce the periodic patterning of the particle distribution. Through these operations, new particles at controlled concentration and flow rate are introduced to the downstream with the surrounding flow field being well-perturbed.}

\begin{figure}[h]
	\centering
	\includegraphics[width=0.62\columnwidth]{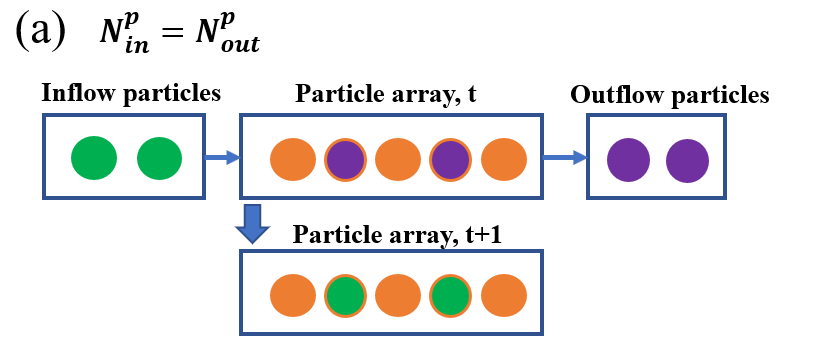}
	\includegraphics[width=0.62\columnwidth]{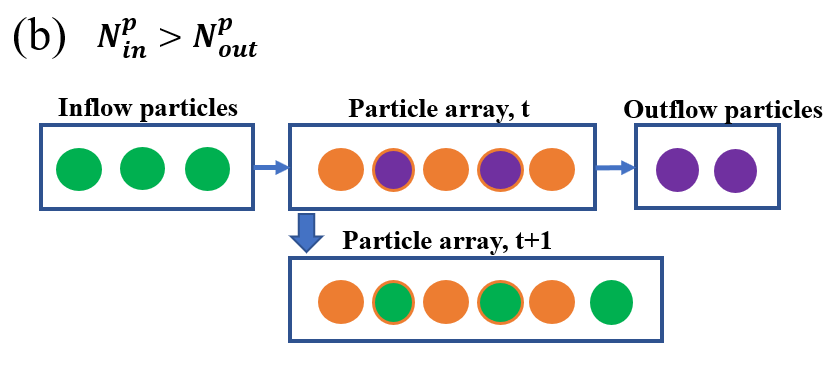}
	\includegraphics[width=0.62\columnwidth]{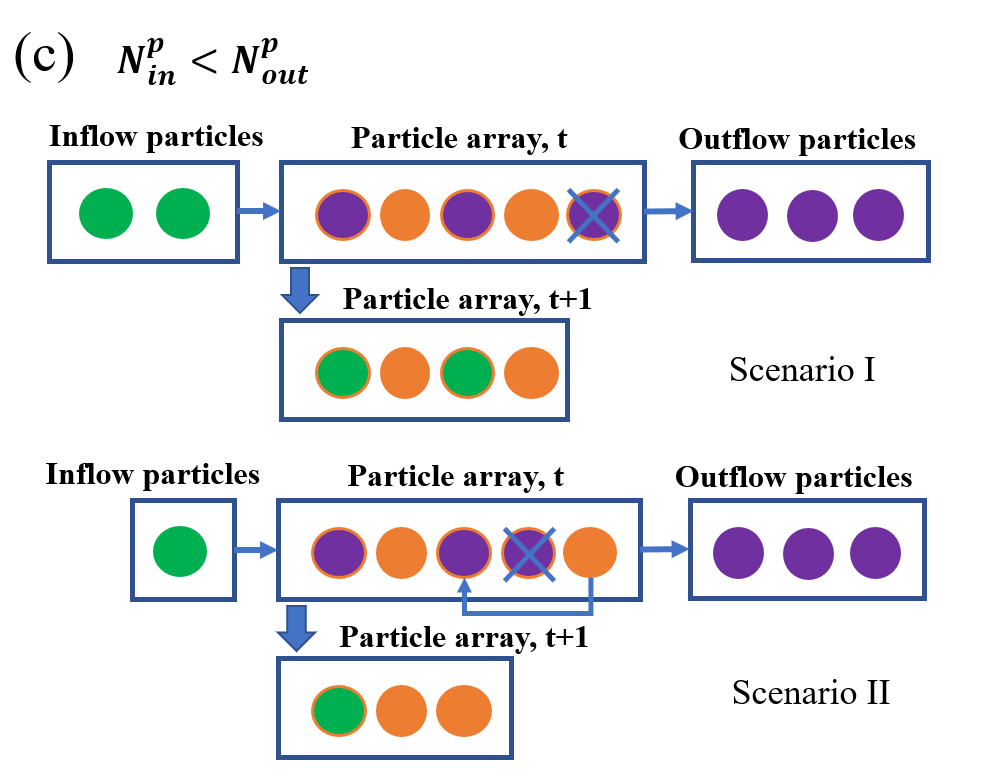}
	\caption{Schematics for the general particulate inflow and outflow boundary condition (PSIO-BC). The particle array represents the array storing the particle information according to the particle index, which increases from 1 to $N_{max}$ from left to right. Here for example $N_{max}=5$ at time $t$
	}
	\label{fig:PSIOalg}
\end{figure}
Similarly, the particle removals are realized by embedding a particle removal region upstream proximal to a single phase region where the outlet fluid BCs are imposed. At the particle removal region, the particle information will be deleted at the local processor (assuming multi-processor parallelization is employed). Meanwhile, the particle index and the total number of removed particles at each time step will be recorded and communicated to the processors in charge of the particle introduction. To minimize the memory allocation for the particle array and perform minimal particle index manipulation, the indices of newly introduced particles are assigned conditionally according to the number of inflow and outflow particles ($N_{in}^p$ and $N_{out}^p$). The schematics for the PSIO-BC operations are shown in Fig~\ref{fig:PSIOalg}.

\begin{algorithm}[H]
	\caption{The Particulate Suspension Inflow and Outflow (PSIO) Algorithm}
	
	{\fontsize{10}{8}\selectfont
		\begin{algorithmic}[1]\label{alg:PSIOBC}
			\REQUIRE $i,j,i_p,j_p,k_p\in \mathbb{Z}^+\cup\{0\}$,
			\STATE Count the number of inflow particles $N_{in}^p$
			\STATE Store properties (location, velocity, etc) of inflow particles $\mathcal{P}_{in}^p(i)|_{i\in\{0,...,N_{in}^p-1\}}$
			\STATE Count the number of outflow particles, $N_{out}^p$
			\STATE Store indices of outflow particles, $\boldsymbol{I}_{out}^p(i)|_{i\in\{0,...,N_{out}^p-1\}}$
			\STATE Update total particle number: $N^p_t \gets N^p_{t-1} + N_{in}^p - N_{out}^p$
			\STATE $\bold{if}$ $N^p_{in} = N^p_{out}$
			\STATE\ \ \ \ $\bold{for}$ $i \gets 0$ to $N_{out}^p-1$
			\STATE\ \ \ \ \ \ \ \ Assign $\mathcal{P}_{in}^p(i)$ to particle $\boldsymbol{I}_{out}^p(i)$
			\STATE\ \ \ \ $\bold{end}\ \bold{for}$
			\STATE $\bold{else}$ $\bold{if}$ $N^p_{in} > N^p_{out}$
			\STATE\ \ \ \ $\bold{for}$ $i \gets 0$ to $N_{out}^p-1$
			\STATE\ \ \ \ \ \ \ \ Assign $\mathcal{P}_{in}^p(i)$ to particle $\boldsymbol{I}_{out}^p(i)$
			\STATE\ \ \ \ $\bold{end}\ \bold{for}$
			\STATE\ \ \ \ Allocate memory for particles $N^p_{t-1}$, ..., $N^p_{t}-1$
			\STATE\ \ \ \ $\bold{for}$ $i \gets N^p_{t-1}$ to $N^p_{t}-1$
			\STATE\ \ \ \ \ \ \ \ $i_p \gets i-N^p_{t-1}+N^p_{out}$
			\STATE\ \ \ \ \ \ \ \ Assign $\mathcal{P}_{in}^p(i_p)$ to particle $i$
			\STATE\ \ \ \ $\bold{end}\ \bold{for}$
			\STATE $\bold{else}$ $\bold{if}$ $N^p_{in} < N^p_{out}$
			\STATE\ \ \ \ $j_p \gets 0$
			\STATE\ \ \ \ $\bold{for}$ $i \gets 0$ to $N_{in}^p-1$
			\STATE\ \ \ \ \ \ \ \ $i_p \gets \boldsymbol{I}_{out}^p(i)$
			\STATE\ \ \ \ \ \ \ \ $\bold{while}$ $i_p\ge N^p_t$
			\STATE\ \ \ \ \ \ \ \ \ \ \ \ $j_p \gets j_p+1$
			\STATE\ \ \ \ \ \ \ \ \ \ \ \ $i_p \gets \boldsymbol{I}_{out}^p(j_p)$
			\STATE\ \ \ \ \ \ \ \ $\bold{end}\ \bold{while}$
			\STATE\ \ \ \ \ \ \ \ $j_p \gets j_p+1$
			\STATE\ \ \ \ \ \ \ \ Assign $\mathcal{P}_{in}^p(i)$ to particle $i_p$
			\STATE\ \ \ \ $\bold{end}\ \bold{for}$
			\STATE\ \ \ \ $k_p \gets 0$
			\STATE\ \ \ \ $\bold{for}$ $i \gets N^p_t$ to $N^p_{t-1}-1$
			\STATE\ \ \ \ \ \ \ \ $\bold{for}$ $j \gets 0$ to $N^p_{out}-1$
			\STATE\ \ \ \ \ \ \ \ \ \ \ \ $\bold{if}$ $i \neq \boldsymbol{I}_{out}^p(j)$
			\STATE\ \ \ \ \ \ \ \ \ \ \ \ \ \ \ \ $\boldsymbol{I}_{s}^p(k_p) \gets i$
			\STATE\ \ \ \ \ \ \ \ \ \ \ \ \ \ \ \ $k_p \gets k_p + 1$
			\STATE\ \ \ \ \ \ \ \ \ \ \ \ $\bold{end}\ \bold{if}$
			\STATE\ \ \ \ \ \ \ \ $\bold{end}\ \bold{for}$
			\STATE\ \ \ \ $\bold{end}\ \bold{for}$
			\STATE\ \ \ \ $\bold{for}$ $i \gets j_p$ to $N^p_{out}-1$
			\STATE\ \ \ \ \ \ \ \ \ $k_p \gets i-j_p$
			\STATE\ \ \ \ \ \ \ \ \ Assign $\mathcal{P}_{in}^p(\boldsymbol{I}_{s}^p(k_p))$ to particle $\boldsymbol{I}_{out}^p(i)$
			\STATE\ \ \ \ \ \ \ \ \ Free the memory of particle $\boldsymbol{I}_{s}^p(k_p)$
			\STATE\ \ \ \ $\bold{end}\ \bold{for}$
			\STATE $\bold{end}\ \bold{if}$
		\end{algorithmic}
	}
\end{algorithm}

Specifically, if $N_{in}^p=N_{out}^p$, the inflow particle properties are assigned to the indices corresponding to the outflow particles, as indicated in Fig~\ref{fig:PSIOalg}a. If $N_{in}^p>N_{out}^p$, in addition to the same procedure as above, the data structure for storing all the particle information in the computational domain (referred to as particle array) will be extended to accomodate for the incremental particles, shown in Fig~\ref{fig:PSIOalg}b. \textcolor{black}{If $N_{in}^p<N_{out}^p$, the particle array will be truncated to reflect the reduction of the total particle number. When the truncated particle indices are all associated with outflow particles (denoted as Scenario I in Fig~\ref{fig:PSIOalg}c), the outflow particles and their memory allocation will be simply eliminated, and the inflow particles will take the indices of the outflow particles starting from the lowest index. When the truncated particle indices contain particles remaining in the computational domain (denoted as Scenario II in Fig~\ref{fig:PSIOalg}c), in addition to the operations in Scenario I, the remained particle information will be assigned to the unoccupied outflow particle indices that are within the range of the updated particle array. The detailed procedure for the PSIO-BC algorithm is listed in the pseudo-code shown in Algorithm~\ref{alg:PSIOBC}.}

\subsection{Reconstruction of physiologic microvascular bifurcations}\label{sec:recon}
Microvascular networks are comprised of many generations of single bifurcations connecting all the way from arterioles or venules to capillaries~\cite{fung2013biomechanics}. Each single bifurcation follows certain physiologic laws that constrain the geometric relation between parent and daughter branches. The radii of parent and daughter branches have been found to obey the Murray's law~\cite{murray1926} stated as,
\begin{equation}
\begin{aligned}
  R_0^3=R_1^3+R_2^3,
  \label{eqn:bifgeo1}
\end{aligned}
\end{equation}
where $R_0$ is the radius of the parent branch; $R_1$ and $R_2$ are the radii of the daughter branches. Complying to this simple geometric relationship leads to a more uniform distribution of vessel wall shear stress and consequently minimizes the power consumption~\cite{sherman1981}. 
The bifurcating angles of two daughter branches also satisfy certain physiologic optimality relating to the vessel radius based on Zamir's law~\cite{zamir1978},
\begin{equation}
\begin{aligned}
  \cos\alpha_1=\frac{R_0^4+R_1^4-R_2^4}{2R_0^2R_1^2}; \ \  \cos\alpha_2=\frac{R_0^4+R_2^4-R_1^4}{2R_0^2R_2^2},
  \label{eqn:bifgeo2}
\end{aligned}
\end{equation}
where $\alpha_1$ and $\alpha_2$ are the bifurcating angles of the two daughter branches with respect to the axial direction of the parent branch. Obeying Zamir's law minimizes the lumen volume of a single bifurcation~\cite{zamir1978}. In addition, we also assume the axes of three branches of a single bifurcation are in-plane as suggested previously~\cite{zamir1983}. The vessel length is assumed to be proportional to the vessel diameter with certain randomness~\cite{lee2018computational}. The cross-section shape of the branches are assumed to be circular.

\begin{figure}[H]
\centering
\includegraphics[width=0.8\columnwidth]{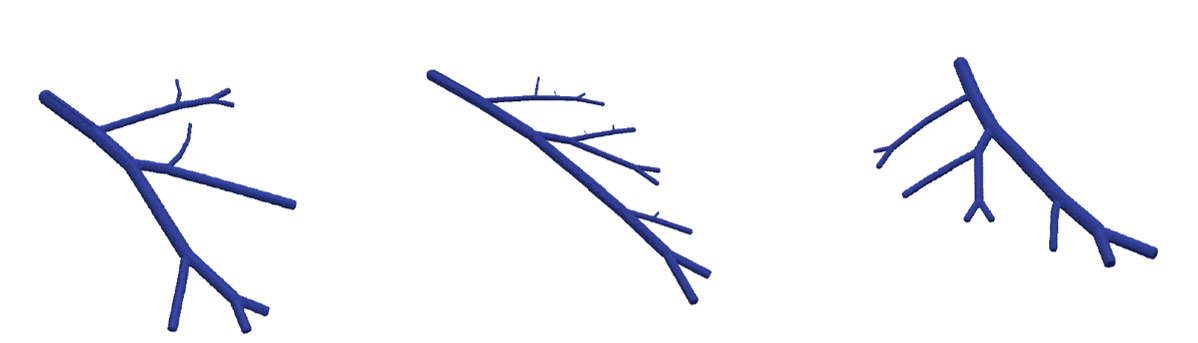}
\caption{Demonstration of three reconstructed microvascular bifurcations of physiologic relevance.}
\label{fig:bifgeo}
\end{figure}

Based on the above physiologic constraints, microvascular bifurcations of multiple generations of branches can be reconstructed as shown in Fig~\ref{fig:bifgeo}, provided the radius of the parental branch, $r_0$, and the size ratio of two daughter branches, $\eta_D=\frac{R_1}{R_2}$, being prescribed randomly within the physiological range~\cite{zamir1983}.

\subsection{Coupling schemes}\label{sec:fsi}
The coupling between fluid and RBC is accomplished through the Aidun-Lu-Ding (ALD) fluid-solid interaction scheme~\cite{Aidun1998,Reasor2012,Aidun2010}. For the suspension dynamics of NP subjected to hydrodynamics and long-ranged many-body HI, the LD particle and the non-fluctuating LB fluid phase are two-way coupled, as discussed and verified in previous studies~\cite{Liu2018a,Liu2018b}.

The RBC-RBC and RBC-vessel wall interactions that specify $\mathbf{f}_n^{CC}$ are based on the subgrid contact functions originally formulated in \citet{DingAidun2003} and later improved by \citet{MacJFM2009} and \citet{ClausenJFM2011}. It prevents the RBC from overlapping when cell-cell/wall separation is below one LB lattice spacing. In this model, the lubrication term is replaced with an exponential contact function to avoid numerical instability driven by the singular nature of the lubrication hydrodynamics and the discrete nature of the interparticle separation calculation, as explained in detail in \cite{MacJFM2009,Clausen2010pof}. 

The NP-RBC contact model that provides $U_{total}$ is based on the Morse potential that can be calibrated to match experimentally measured surface interaction forces \citep{Neu2002,Liu2004}. Due to the lack of statistics for actual NP-RBC short-distance interaction, this study employs the measured cell-cell interaction potential \citep{Neu2002} for the NP-RBC interactions. The model parameters are specified according to \cite{Liu2004} but with a cut-off distance to exclude the attraction. 
The short-distance NP-NP interaction is neglected due to the extreme dilution of the NP concentration ($\ll$1$\%$) considered.

\section{In vivo experiment}\label{sec:exp}
To validate the numerical methods, the blood flow through a microvascular bifurcation within the chicken chorioallantoic membrane (CAM) in vivo model has been characterized through microscale particle image velocimetry ($\mu$PIV). The CAM model is a highly vascularized extraembryonic membrane that is readily accessible for intravital imaging. In-house synthesized mesoporous silica nanoparticles (MS-NP) were used as tracking particles. 

\begin{figure}[H]
\centering
\includegraphics[width=0.98\columnwidth]{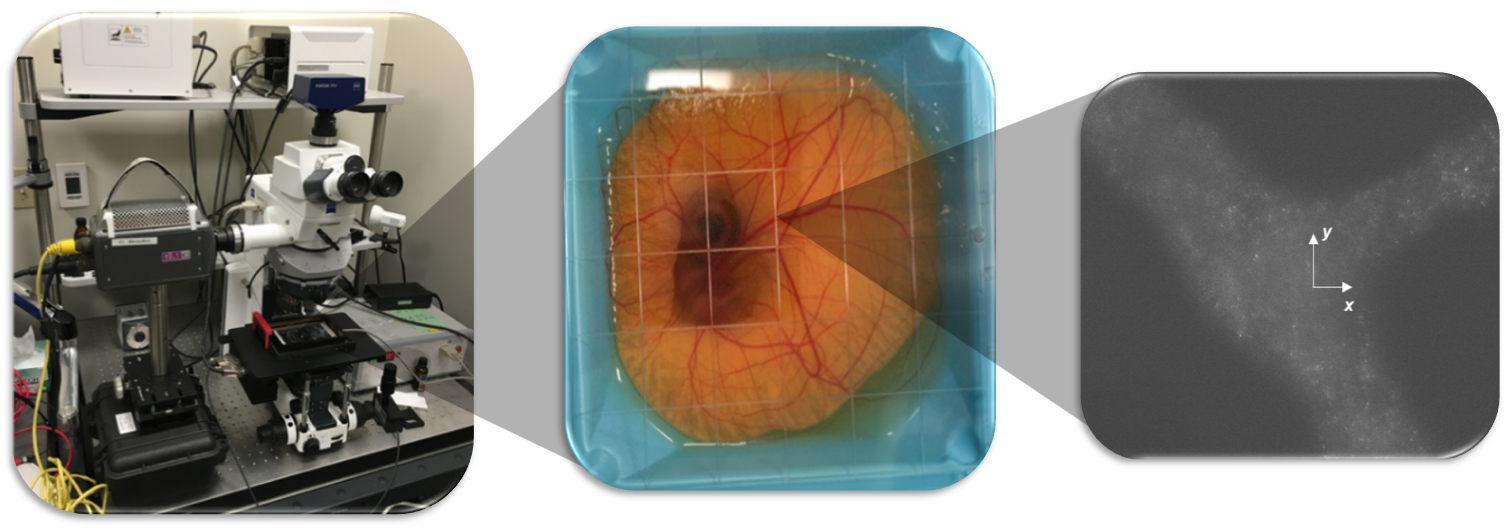}
\caption{The benchtop microscale particle image velocimetry ($\mu$PIV) system (left), the chick chorioallantoic membrane microvasculature model (middle) and the grey-scale visualization of the flow field of a selected region of interest (right).}
\label{fig:exp}
\end{figure}

The $\mu$PIV system is shown in Fig~\ref{fig:exp}~(left), which mainly consists an upright microscope (Zeiss Axio Examiner Z.1) modified with a heated stage operated at 63 magnification to image the CAM model (10 days post fertilization). The microscope is fitted with a water immersion lens (LDC-APO Chromat 63 1.15W) to facilitate imaging of deep vessels. Imaging was performed using a high-speed camera (Hamamatsu Orca Flash 4.0) operated at 142 frames/sec, enabling a resolution of 0.1 microns/pixel. A broadband light source was used along with careful image and PIV processing methods through the LaVision PIV software package DaVis v8.4 to mitigate image blur effects especially at peak velocities. 

The CAM model is depicted in Fig~\ref{fig:exp}~(middle). The ex-ovo avian embryos were handled according to published methods~\cite{leong2010intravital} with all experiments conducted following an institutional approval protocol (11-100652-T-HSC). Embryos were utilized for experimentation 7 days post removal from shells (day 10 post fertilization). A dosage of 50 $\mu$g MS-NPs with size $\sim$130 nm were injected into tertiary veins of the CAM. After injection, embryos were placed in a customized avian embryo chamber for imaging.

Applying the preceding methodology, a greyscale image of a vascular bifurcation is shown in Fig~\ref{fig:exp}~(right). The image location was chosen such that the vessel branches remain relatively flat over the entire field-of-view. To obtain the dynamic velocity field, 5000 images were acquired in total. A sliding sum-of-correlation method was used to calculate vector fields~\cite{lee2007micro}. In \S~\ref{sec:vef2}, the mean velocity field obtained based on the NP flow is compared with the computed counterparts through the current multiscale computational framework.

\section{Results and Discussion}\label{sec:results}
In this section, we first verify the PSIO-BC by recovering a tubular blood flow and comparing the results with that obtained from the typical periodic boundary condition. Furthermore, the multiscale computational framework is further validated by reconstructing the CAM bifurcation flow in silico and comparing simulation against the experimental measurement. As follows, the NP partitioning pattern in the presence of RBCs is interrogated quantitatively by simulating a bidisperse RBC-NP suspension through single physiologically-relevant microvascular bifurcation under different flow partition ratios with or without geometric asymmetry. Physical mechanisms leading to specific NP partitioning characteristics are discussed in detail.

\subsection{Recovering tubular flows}\label{sec:vef1}
\begin{figure}[H]
\centering
\includegraphics[width=0.82\columnwidth]{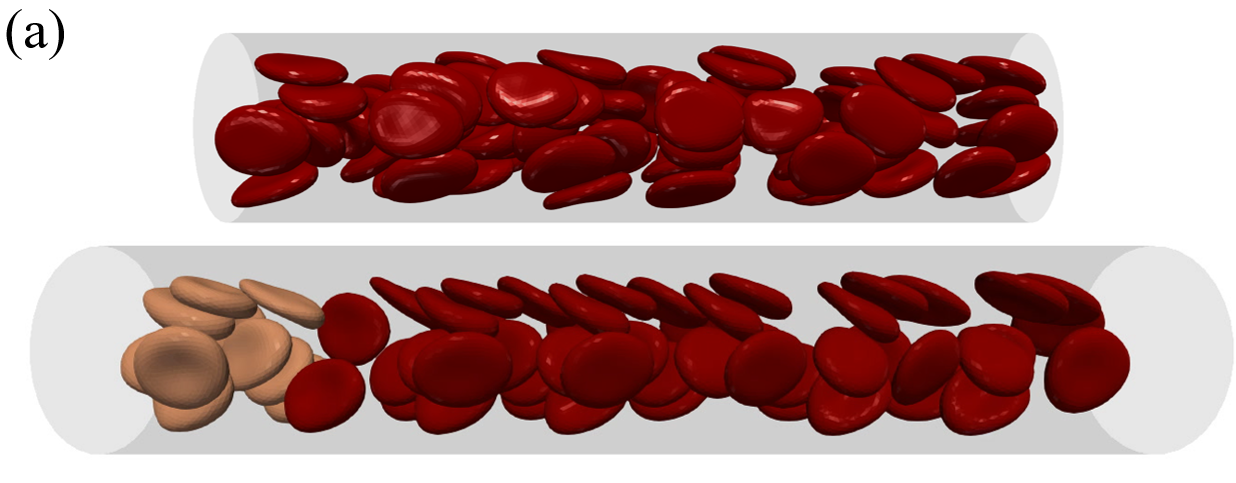}
\includegraphics[width=0.82\columnwidth]{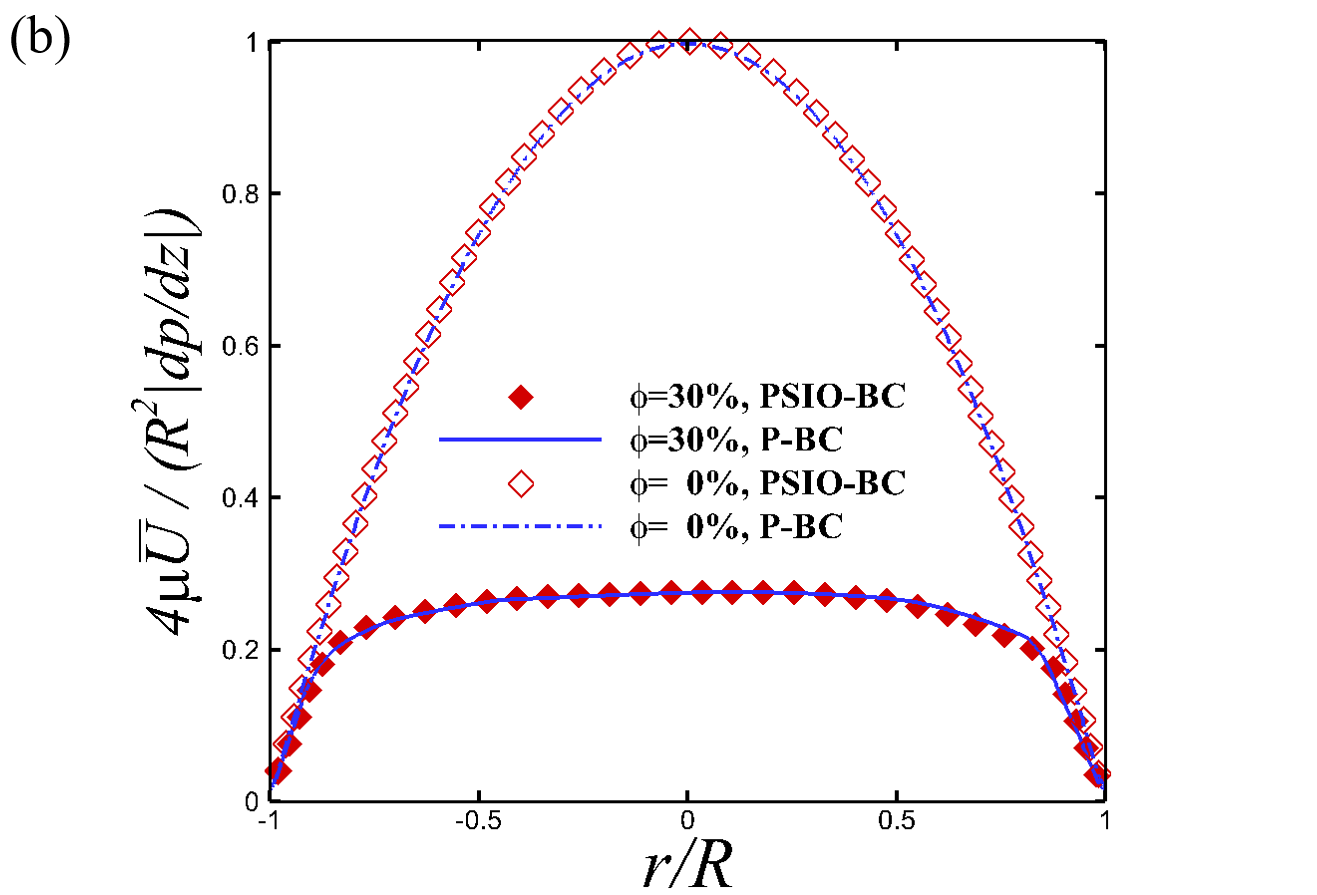}
\caption{Normalized axial flow velocity plotted against vessel radial locations for a tubular flow with or without red blood cells, computed with the P-BC or the PSIO-BC.}
\label{fig:verif}
\end{figure}

Flow through a straight tube with or without RBCs is simulated with both PSIO-BC and P-BC. The tube has a diameter of $2R$=20 $\mu m$ and a length of $L$=80 $\mu m$. The fluid viscosity is selected as $\mu$=1.2 cP. A hematocrit of $\phi$=30\% is selected for the multiphase simulations. The RBC has a shear modulus of $G$=0.0063 $dynes/cm$. For the P-BC simulations, a constant pressure gradient, $-\frac{dp}{dz}$=240 $Pa/mm$, is applied to drive the flow (from left to right according to Fig~\ref{fig:verif}a) such that the wall shear rate is $\dot{\gamma}_w$=1000 $s^{-1}$. For the single phase simulation using the PSIO-BC, parabolic velocity profiles, $u(r)$=$|\frac{dp}{dz}|\frac{(R^2-r^2)}{4\mu}$, are prescribed at both the inlet and outlet. For RBC case through PSIO-BC, inlet and outlet pressures are prescribed at the outmost ends of the single phase regions. The inlet pressure is $p_i$=$p_0$+$|\frac{dp}{dz}|(L+L_{BC})$ and the outlet pressure is $p_o$=$p_0$, where $p_0$ is chosen as the atmospheric pressure and $L_{BC}$=60 $\mu m$ that accounts for the extra tube sections for PSIO-BC.
 
The velocity profiles for the tubular flow simulations using both PSIO-BC and P-BC are plotted in Fig~\ref{fig:verif}b. The velocity profiles for the cases with $\phi$=30\% are averaged over 10 different axial locations at one time point during equilibrium. It is shown that the results via PSIO-BC agrees well with that via P-BC, showing the validity of the PSIO-BC algorithm.

\subsection{Comparing with in vivo measurements}\label{sec:vef2}
\begin{figure}[H]
\centering
\includegraphics[width=0.85\columnwidth]{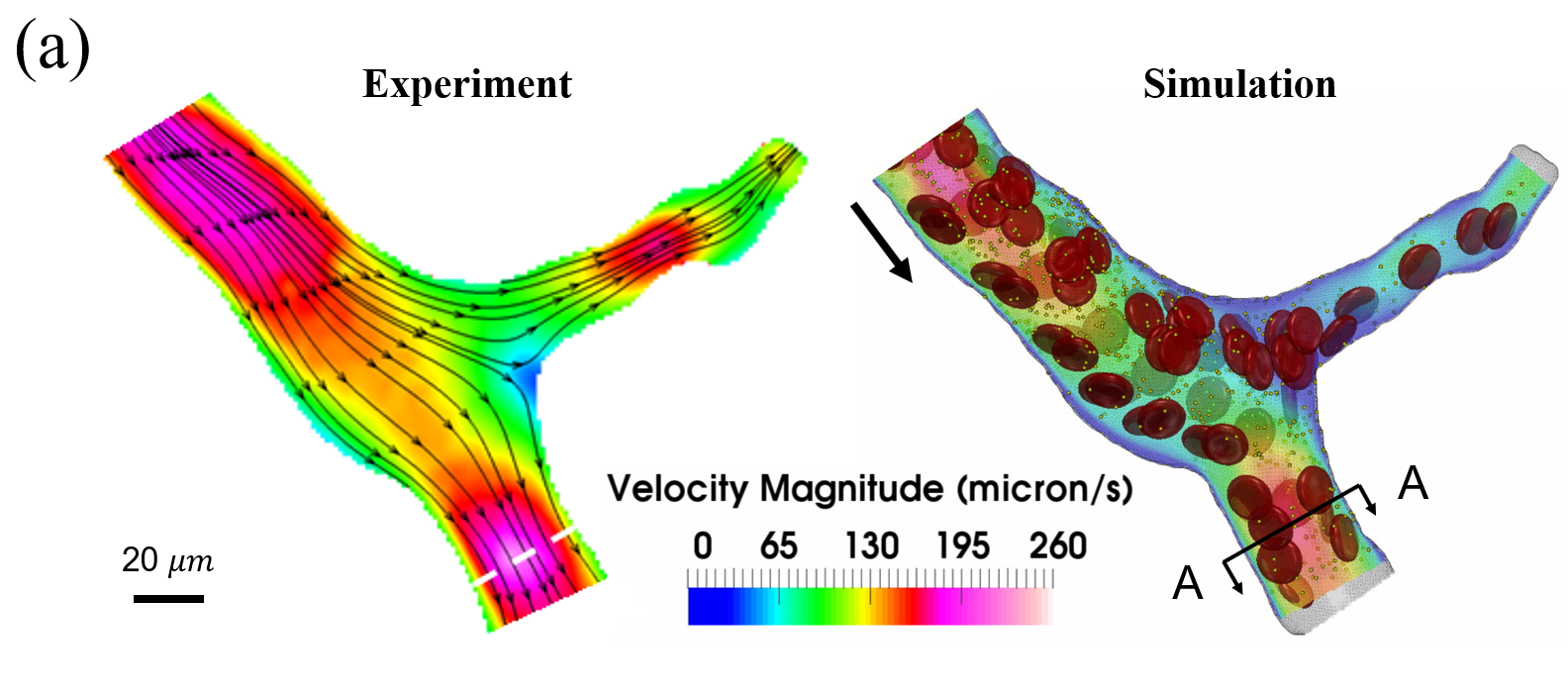}
\includegraphics[width=0.85\columnwidth]{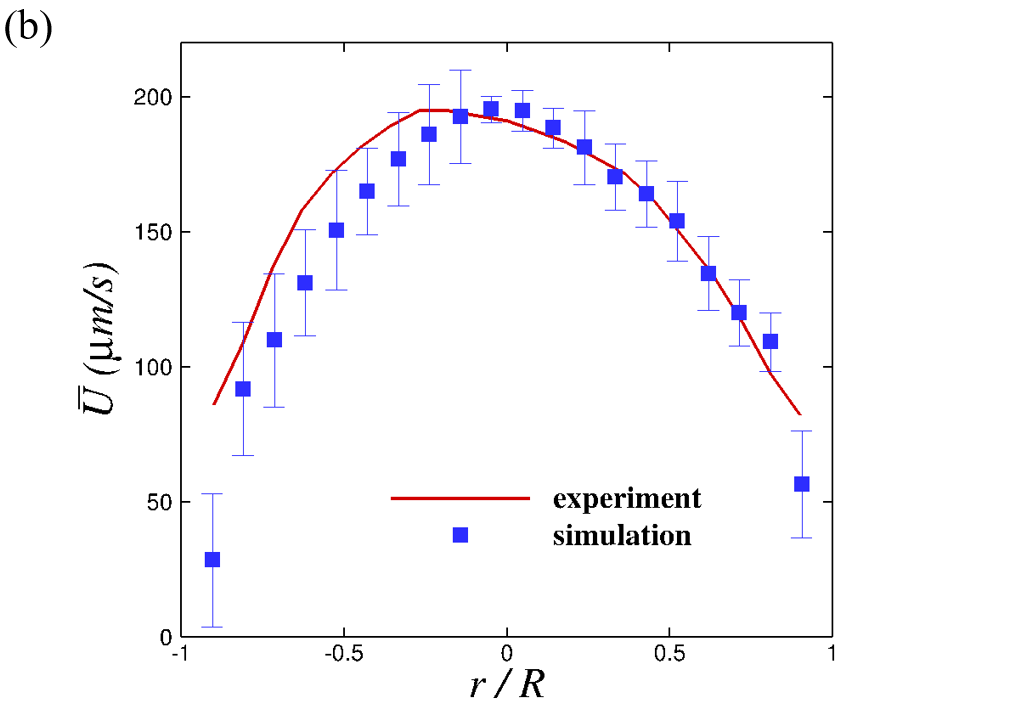}
\caption{(a) Comparisons of the mean velocity distribution between the experiment and simulation through a CAM vascular bifurcation. The velocity contour for the simulation is an snapshot after the system reaches equilibrium (i.e., the number of particles reach plateau).  (b) The radial distribution of the mean axial velocity at the A-A cross-section annotated in (a). The asymmetry of the experimental curve may be related to the non-circular shape of the actual cross-section area.}
\label{fig:valid}
\end{figure}

To further validate the current multiscale computational framework, we simulate a NP-RBC suspension flow through a CAM microvascular bifurcation (see supplemental video 1) and compared the results against experimental measurements through $\mu$PIV. An arterial branch in the CAM model is selected for the study due to its relatively big size and flat orientation, which are suitable for measurement. The parent branch has a radius of $R_0$=20 $\mu m$ and the two daughter branches have radii of $R_1$=7.5 $\mu m$ and $R_2$=17.5 $\mu m$ respectively, shown in Fig~\ref{fig:valid}a. Slight pulsatility of the flow is observed with a frequency of $f=3\ Hz$. The Womersley number for this problem can be estimated as $\alpha$=$\sqrt{\rho fR^2_0/\mu}\approx0.03\ll1$, which suggests the pulsation effect is insignificant. Therefore, the study below concentrates on comparing the mean velocity as a reasonable simplification.

Contour maps of the mean velocity field from the $\mu$PIV measurement are demonstrated in Fig~\ref{fig:valid}a on the left. The mean streamlines are overlaid, indicating the flow direction. We simulate the NP-RBC suspension flow through the same reconstructed geometry using the multiscale computational method, where Dirichlet velocity BCs based on the in vivo measurements are applied at the inlet and outlet using the PSIO-BC. The hematocrit is assumed to be 20\% which is an average physiological value for CAM blood flows~\cite{maibier2016structure}. The NP number concentration is set to $10^7/ml$ for statistic purposes. The RBC size is selected to be the same as human RBC size, while the rigidity (shear modulus) of the RBC membrane is elevated by three folds to be consistence with low deformability observed in avian RBCs~\cite{gaehtgens1981comparative}. As shown in Fig~\ref{fig:valid}, the fluid velocity distribution reproduced by simulation qualitatively matches well with the PIV results.

The velocity profiles across the diameter of the larger daughter branch at the A-A location (denoted) are further plotted in Fig~\ref{fig:valid}b. The velocity profile in the simulation is obtained by tracking the streamwise velocity of the NP flowing through the A-A cross-section at equilibrium to mimic the $\mu$PIV process. The mean velocity near the wall has a finite number for both the simulation and experiment as a result of the sliding nature of the NP adjacent to the wall. The velocity profile in the RBC-laden region shows overall good agreements between experiment and simulation. 

\subsection{NP partition through microvascular bifurcations}\label{sec:actualres}
To understand how NPs partition at the junction of a microvascular bifurcation, a trade study of NP-RBC suspension flow through single microvascular bifurcations is performed, taking into account the asymmetry in both the flow partitioning and vascular geometry. To control the partitioning of the bulk fluid, a flow partition ratio (FPR) is defined as $\eta_{Q,i}^F = Q_i/Q_0$ where $Q_i$ ($i$=1 or 2) denotes the suspension flow rate at one daughter branch and $Q_0$ is the suspension flow rate at the parental branch. Similarly, a particle partition ratio (PPR) is defined as $\eta_{Q,i}^p = N_i/N_0$, where $N$ denotes the number of particles entering the corresponding branch. The PPR is used to quantitatively describe the partition pattern of the particle phase (RBC and NP) under different FPR. For the following studies, the parental branch has a size of $2R_0$=20 $\mu m$, a wall shear rate of $\dot{\gamma}$=2000 $s^{-1}$ and a hematocrit of $\phi$=20\%, reflecting physiologic conditions of arterioles~\cite{pries2008blood,fung2013biomechanics}. The Dirichlet velocity boundaries are applied to both the inlet and outlet, where the Poiseuille flow velocity profiles are imposed at the single phase region pertaining to specific $\eta_{Q,i}^F$. The vascular geometric effect is controlled by the daughter branch size ratio, $\eta_D$. Provided $R_0$ and $\eta_D$, the exact size and bifurcating angle of the two daugher branches are uniquely constrained by the physiologic laws discussed in~\ref{sec:recon}.

\subsubsection{Symmetric bifurcations}
\begin{figure}[H]
	\centering
	\includegraphics[width=0.8\columnwidth]{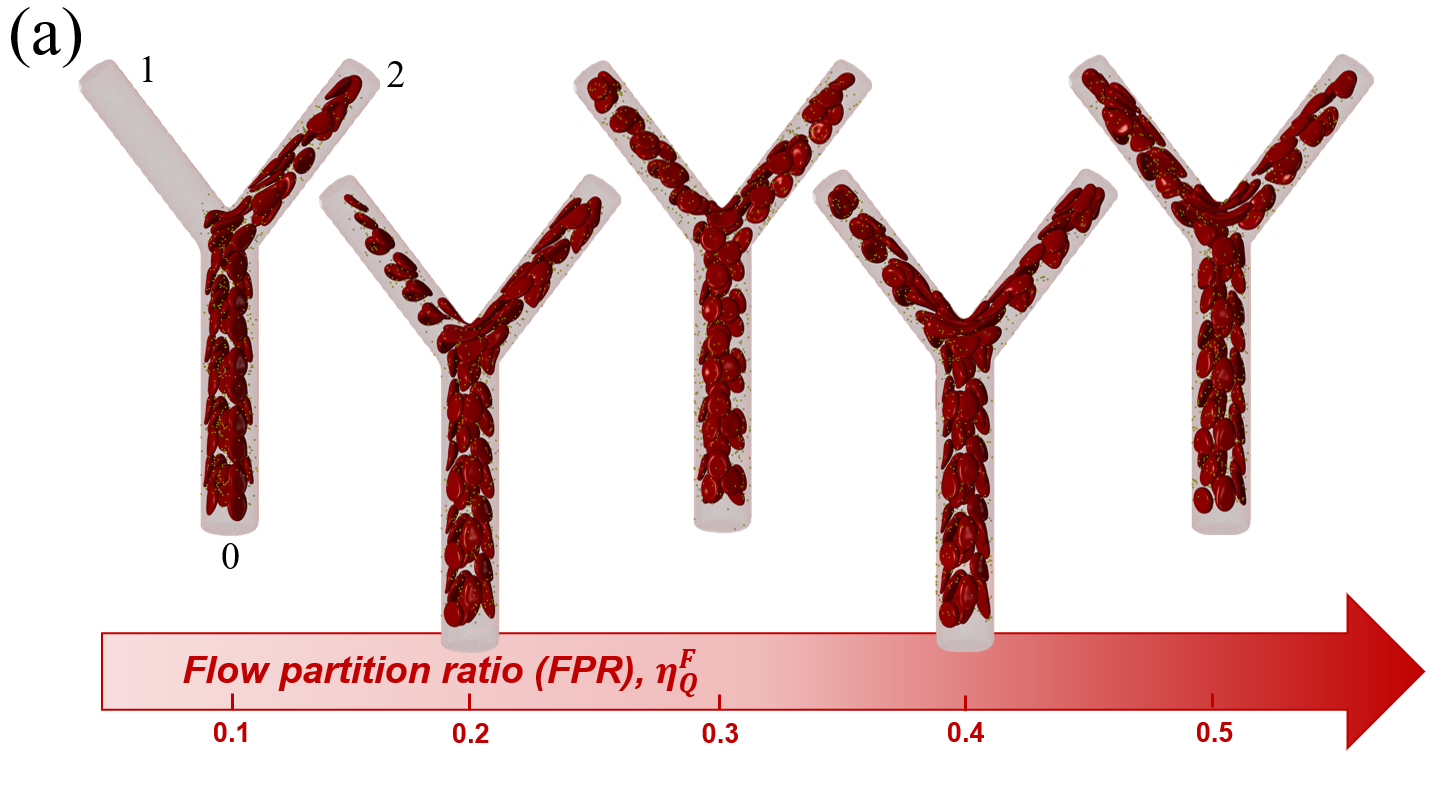}\\
	\includegraphics[width=0.8\columnwidth]{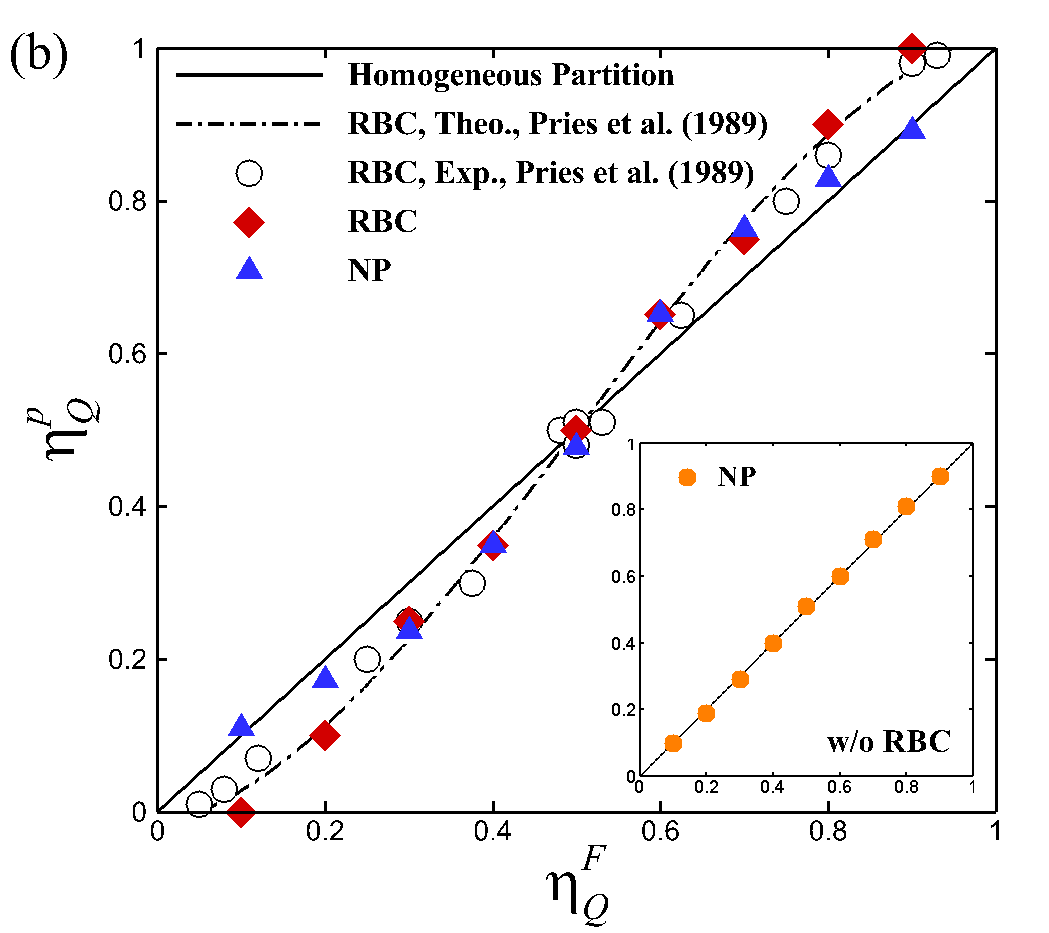}
	\caption{(a) Snapshots of RBC-NP suspension flow through symmetric bifurcations under various suspension flow flux partition ratio, $\eta_Q^F$. (b) Particle flow partition ratios, $\eta_Q^p$, plotted against suspension flow partition ratio, $\eta_Q^F$, for symmetric microvascular bifurcations. The inset shows the results without the presence of RBCs.}
	\label{fig:sym2}
\end{figure}

We first study the NP-RBC suspension flow partitioning through a geometrically symmetric bifurcation by setting $\eta_D=1$, as shown in Fig~\ref{fig:sym2}a. 
As a control, when there is no RBC in the bifurcation, the $\eta^p_{Q}$ of NP changes linearly as the $\eta^F_{Q}$ changing from 0.1 to 0.9, as shown in the inset of Fig~\ref{fig:sym2}b. This partitioning pattern is termed homogeneous partition, as the flow of the particle phase follows the fluid phase while branching towards the downstream.

When RBCs are introduced from the parental branch, the simulated $\eta^p_{Q}$ of the RBC phase exhibits a sigmodal trend with respect to $\eta^F_{Q}$, as shown in Fig~\ref{fig:sym2}b. This trend agrees well with both the experimental measurement and theoretical correlation by \citet{pries1989red}, indicating the classic ZF effect is well captured by the current computational methods. Such heterogeneous partition of RBC is most prominent at $\eta^F_{Q}=0.1$ or 0.9, where complete phase separation of RBC occurs in one of the daughter branches, as shown in Fig~\ref{fig:sym2}a. Similar to the ZF effect in the RBC phase, the $\eta^p_{Q}$ for NP also deviates from the linear line, suggesting the NP partitioning also becomes heterogeneous. However, such heterogeneity in the NP partitioning seems to be conditional as the $\eta^p_{Q}$ falls back on the linear line at about $\eta^F_{Q}$=0.1 or 0.9 while complete phase separation of RBC occurs.

\subsubsection{Asymmetric bifurcations}
\begin{figure}[H]
	\centering
	\includegraphics[width=0.8\columnwidth]{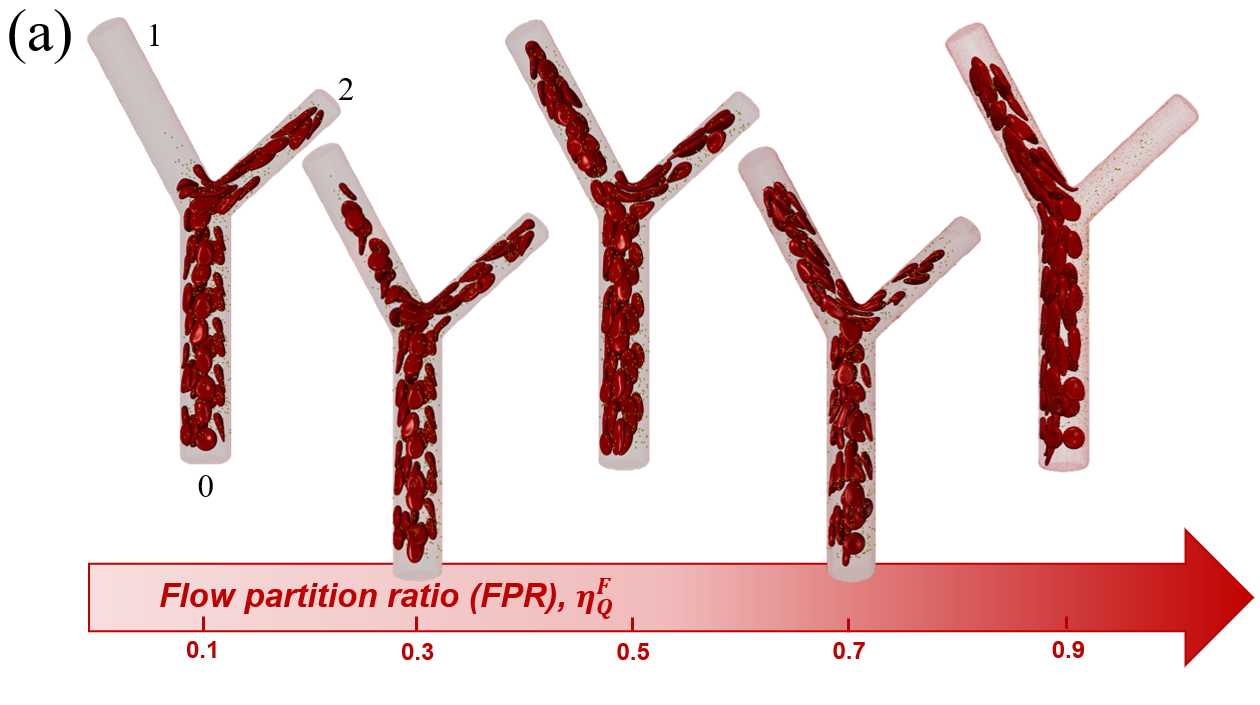}\\
	\includegraphics[width=0.8\columnwidth]{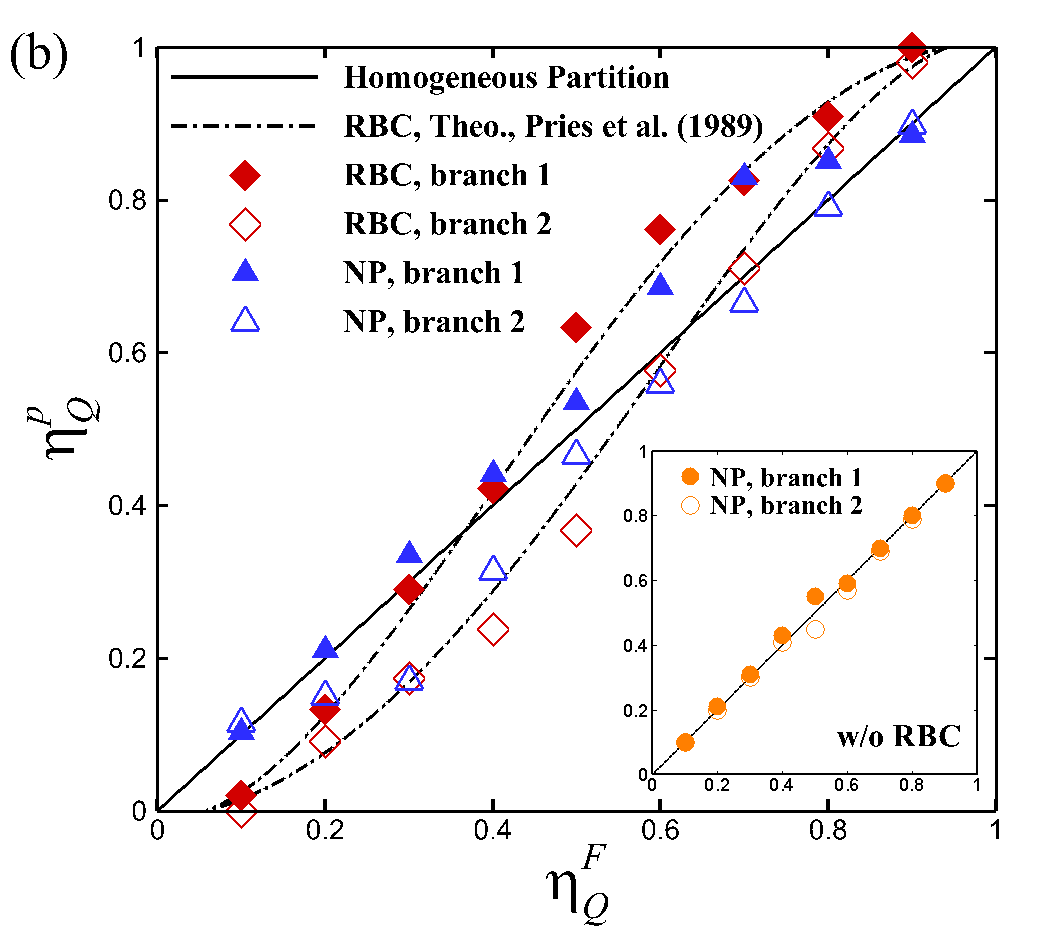}
	\caption{(a) Snapshots of RBC-NP suspension flow through asymmetric bifurcations under various suspension flow flux partition ratio, $\eta_Q^F$. (b) Particle partition ratio, $\eta_Q^p$, plotted against suspension flow partition ratio, $\eta_Q^F$, for asymmetric microvascular bifurcations. The inset shows the results without the presence of RBCs.}
	\label{fig:asym2}
\end{figure}

To introduce the effect of geometric asymmetry, we set $\eta_D=0.8$ as a physiological ratio of asymmetry reported in \citet{zamir1983}. The resulting geometry of the asymmetric bifurcation is shown in Fig~\ref{fig:asym2}a.
Similar to the symmetric bifurcation case, NP shows mainly homogeneous partition when RBCs are not present, as shown in the inset of Fig~\ref{fig:asym2}b. However, due to geometric asymmetry, the $\eta_Q^p$ of NP is deviated from the linear curve at $\eta_Q^F=0.5$. This can be explained by the slight deviation of the bifurcation junction towards the smaller daughter branch which causes the uniformly dispersed NPs \cite{Liu2019b} prioritizing entering the larger daughter branch (branch 1). 

As further shown in Fig~\ref{fig:asym2}b, the asymmetric bifurcation exhibits a more pronounced heterogeneity in the RBC partition, as predicted by both our simulations and the theory in \citet{pries1989red}. Different from the symmetric bifurcation, the heterogeneity of RBC partition even occurs during equal flow partitioning ($\eta_Q^F=0.5$) as another manifestation of the effect of the geometric asymmetry. The enhanced heterogeneity of the RBC partitioning further elevates the heterogeneity in the NP partitioning compared to the symmetric bifurcation case. Again, the partitioning of NP recovers homogeneity at extreme flow partition ratios when severe phase separate occurs. 

\subsection{Mechanisms for conditional heterogeneity in NP partition}
Next, we take a detailed look at the particle trajectory and flow structure of the asymmetric bifurcation case to understand the underlying mechanisms governing the conditional heterogeneity in the NP partition through microvascular bifurcations. Specifically, we look at two flow partition ratios, $\eta_Q^F=0.7$ and 0.9, that gives rise to the high heterogeneity in NP partition and the recovery of homogeneous partition, respectively.

\begin{figure}[H]
\centering
\includegraphics[width=0.95\columnwidth]{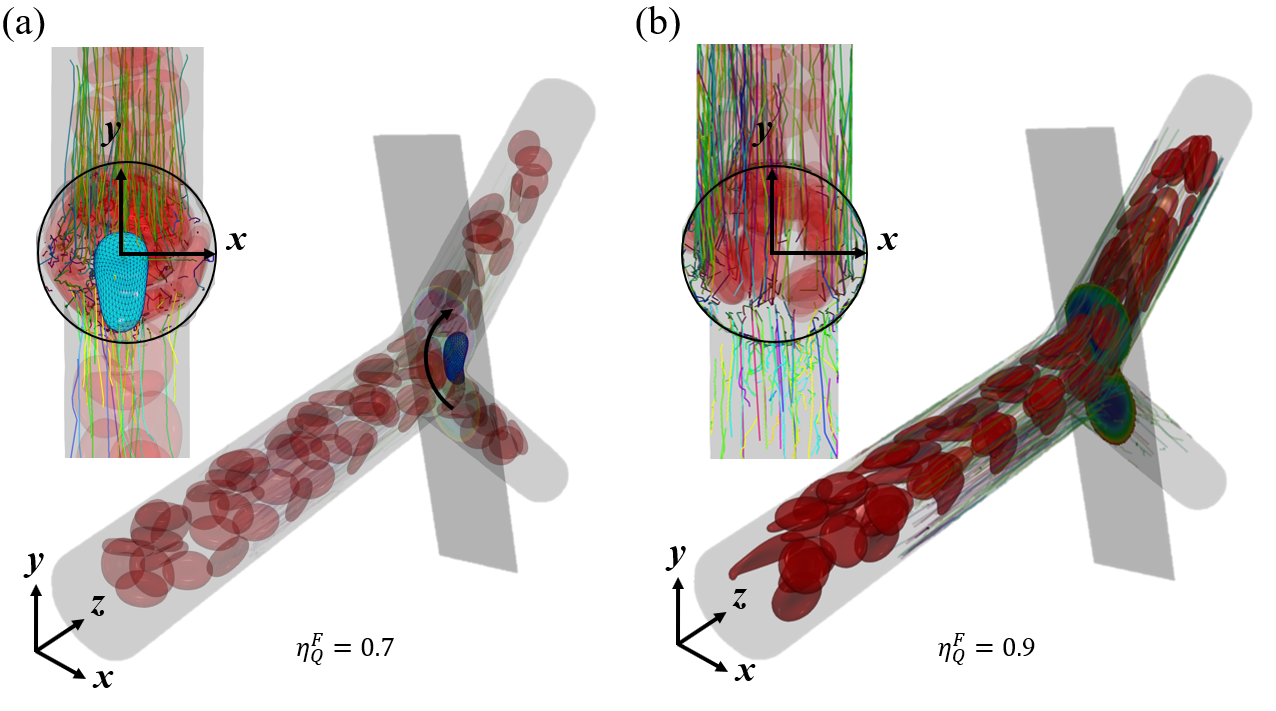}
\includegraphics[width=0.95\columnwidth]{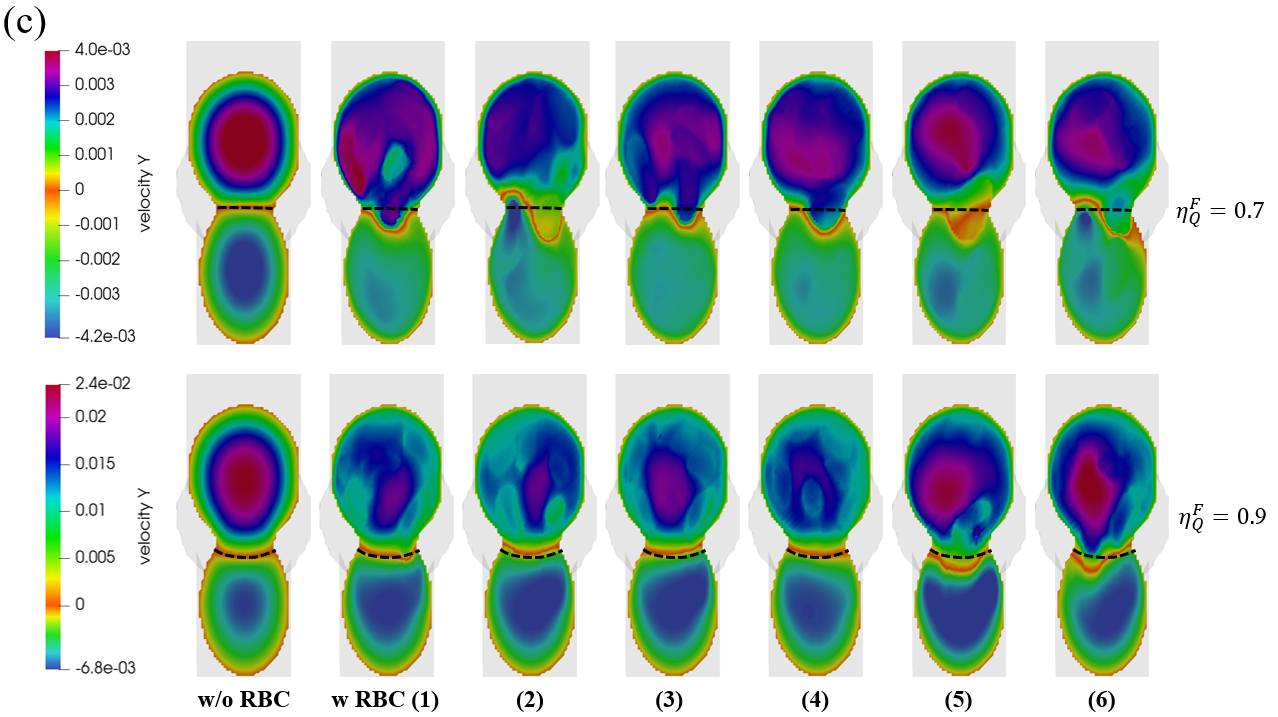}
\caption{(a-b) The cell distribution, NP trajectories and a cross-section of interest within an asymmetric microvascular bifurcation. (c) The contours of the Y velocity component at the cross-section adjacent to the bifurcation junction. The left column shows the case without the presence of RBCs. The other six columns corresponds to six time points when a RBC sliding towards the upper branch. The black dashed line denotes the zero velocity line, which shows the flow separatrix that divides the flow entering two daughter branches.}
\label{fig:mbif}
\end{figure}

At $\eta_Q^F=0.7$ , RBCs periodically collide onto the junction, linger for a while and exhibit a higher frequency of entering the higher-velocity branch in consistence with the ZF effect~\cite{fung1973}. As an example shown in Fig~\ref{fig:mbif}a, the RBC (denoted in blue) is lingering while performing a tank-treading motion and sliding into the upper branch where the flow rate is higher. The upper sliding motion occurs more frequently than the lower sliding motion, which consequently leads to $\eta_Q^p=0.83$ being higher than $\eta_Q^F$. The RBC tank-treading motion at the bifurcation junction will further alter the local flow field. As shown in Fig~\ref{fig:mbif}c, at the condition of $\eta_Q^F=0.7$, without RBCs the flow separatrix is well aligned with the geometric dividing line at the junction; with RBCs lingering and tank-treading towards the upper branch, the flow separatrix tends to locally shift towards the lower branch, which locally leads to the recruitment of more NPs towards the upper branch. This RBC-induced entrainment effect seems to be the mechanism that causes the heterogeneity in the NP partition when RBCs are present.

When the recovery of homogeneous partitioning occurs at $\eta_Q^F=0.9$, a certain amount of NPs can still flow into the RBC-free branch, as shown in Fig~\ref{fig:mbif}b. As indicated by the trajectories of NP throughout the bifurcation, the majority of the NPs that enter the lower branch come from the cell-free layer (CFL) of the parental branch. Unlike the deformable RBCs that tend to concentrate to the center of the tube as a result of the lift force exerted by the vessel wall~\cite{abkarian2002tank}, NPs tend to disperse throughout all the radial locations due to the severe Brownian effect~\cite{Liu2018a,Liu2019b}. In other words, the NP phase can always enter the cell-free branch so long as the plasma skimming occurs, which will reasonably contribute to the homogeneity in the NP partitioning. As further shown in Fig~\ref{fig:mbif}b, the flow separatrix at $\eta_Q^F=0.9$ does not seem to be drastically distorted when the RBC is present. Therefore, given the incoming NPs are uniformed distributed in the parental tube~\cite{Liu2019b}, the ratio of NP partitioning with RBCs is expected to be similar to the case without RBCs. 

Taken together, the heterogeneity in the NP partition in the presence of RBCs seems to be associated with an RBC-induced entrainment effect, that distorts the flow separatrix to the low flow rate branch and thus entrains more NPs to the high flow rate branch. The presence of NPs in the CFL accompanied by the normalization of flow separatrix at the extreme flow partition ratios (when blood plasma skimming occurs) leads to the return of NP partitioning from heterogeneity to homogeneity in the presence of RBCs. The competition between th RBC-induced entrainment effect and the presence of NP in the CFL contributes to the conditional heterogeneity of the NP partitioning as observed in~\ref{sec:actualres}.

\section{Summary}\label{sec:conclusion}
In this work, we have developed a multiscale computational framework to directly simulate the NP distribution and partition through the microvascular bifurcations in the presence of RBCs. A general particulate suspension inflow and outflow boundary condition has been developed and embedded in the computational framework, and later applied to both the RBC and NP suspension dynamics. To validate the model, in vivo measurements of the NP flow through a CAM microvascular bifurcation have been performed. The current computational framework is able to quantitatively reconstruct the detailed physiological bifurcation flow by directly simulating both the RBC and NP suspensions. The velocity distribution for the NP phase through simulation matches well with the experimental measurement, showing the accuracy and validity of the present multiscale computational framework. 

Applying the multiscale computational framework, we further interrogate the RBC-NP suspension flow through idealized single bifurcations under physiologically relevant geometry and flow conditions. The classic Zweifach-Fung effect is found to be well captured by the method. Moreover, we find the partition of the blood-borne NP through microvascular bifurcations in response to the ZF effect also shows a substantial heterogeneity. The heterogeneity in the NP partition is however conditional depending on the level of disproportionality in the flow partitioning between the daughter branches. The physical mechanisms responsible for the heterogeneity in the NP partition has been analyzed in terms of the flow structure and the particle trajectories. It is found that the heterogeneity in the NP partition in the presence of RBCs seems to be associated with the flow separatrix deviation at the bifurcation junction caused by the RBC tank-treading motion. The recovery of homogeneity in the NP partitioning however has to do with the plasma skimming of the NPs in the CFL accompanied with the normalization of flow separatrix at the extreme flow partition.

This study presents a numerical tool that can be used to directly analyze the blood-borne NP distribution within microcirculation that often comprises complex branching structures (see the supplemental video 2). The results imply the importance of the presence of the RBC phase in directing the blood solute distributions and provides possible physical mechanisms that contribute to the heterogeneous distribution of blood solute in microvasculatures observed clinically~\cite{jain2010delivering}. Such insights can provide valuable information for a upscaling to a network model of the full vasculature~\cite{lee2018computational} and may help the design of nanocarriers with improved drug delivery efficacy~\cite{wilhelm2016analysis}. It is also noteworthy that the proposed particulate suspension inflow and outflow boundary condition is general and can be applied to other applications such as modeling arterial thrombosis~\cite{kim2019occlusive} or other reacting flows that often involve consumption of particles or species.

\section*{acknowledgements}
The authors acknowledge the partial funding from Sandia National Laboratories made available through contract number 2506X36. The computational resource granted by XSEDE under contract number TG-CT100012 is acknowledged. This work is supported by the Laboratory Directed Research Development Program at Sandia National Laboratories. Sandia National Laboratories is a multimission laboratory managed and operated by National Technology and Engineering Solutions of Sandia LLC, a wholly owned subsidiary of Honeywell International Inc. for the U.S. Department of Energy’s National Nuclear Security Administration under contract DE-NA0003525. This paper describes objective technical results and analysis. Any subjective views or opinions that might be expressed in the paper do not necessarily represent the views of the U.S. Department of Energy or the United States Government.

\bibliography{PRE}

\end{document}